\documentstyle[prl,multicol,aps]{revtex}
\begin{document}

\title
{Thermodynamics of trapped interacting bosons in one dimension} 
\author{Shi-Jian Gu$^1$, You-Quan Li$^{1,2}$ and Zu-Jian Ying$^1$}
\address{
$^1$Zhejiang Institute of Modern Physics,
Zhejiang University, Hangzhou,310027, China\\
$^2$Institute for Physics, Augsburg University, D-86135 Augsburg, Germany
}

\date{Received: }
\maketitle
\begin{abstract}

On the basis of Bethe ansatz solution of bosons with $\delta$-function
interaction in a one-dimensional potential well,
the thermodynamics
equilibrium of the system in finite temperature is studied by using the
strategy of Yang and Yang. The thermodynamics quantities, 
such as specific
heat etc. are obtained for the cases of strong coupling limit and
weak coupling limit respectively. 
\end{abstract}


\section{Introduction}

There has been much interest recently in the study of strongly 
correlated electronic systems in one dimension. 
This is because not only various non-perturbative methods
\cite{Faddeev,Yang,Wiegmann,Andrei,Haldane1,deVega,FrahmKorepin,%
KorepinBook,Schlottmann,Carmelo} are applicable to models 
in one dimension but also several photo-emission 
experiments for one-dimensional compounds of alkali-metal cupper 
oxides\cite{KimShen,Kobayashi,Fujisawa} 
are carried out. Bosons with $\delta$-function interaction 
in one dimension is a simple
but interesting model. It was solved in ref.\cite{LiebL} under the 
periodic boundary condition, in ref.\cite{Gaudin,Woynarovich} 
under the boundary condition  
of potential well of infinite depth
and in ref.\cite{Li} under that of potential well of finite depth.
A strategy for studying the thermodynamics  
of exactly
solvable models was proposed in ref.\cite{YYang}
when discussing the solution of ref.\cite{LiebL}.
In the present paper, using the strategy of ref.\cite{YYang}
we study the thermodynamics  on the basis of 
the Bethe ansatz solution of ref.\cite{Li}.
After recall the model Hamiltonian and the Bethe-ansatz equation we 
study the thermal equilibrium in Section II. The formal expressions
of free energy and pressure are obtained. In Section III we consider
the strong coupling limit and obtain the quasi-momentum distribution 
and specific heat explicitly. 
In Section IV the case of week coupling limit is discussed extensively. 

\section{Thermal equilibrium at finite temperature}

The Hamiltonian of bosons in a one-dimensional potential well 
of finite depth with $\delta$-interaction reads 
\begin{equation}
H=-\sum_{i=1}^{N}\frac{\partial^2}{\partial{x_i}^2}
   +\sum_{i=1}^{N}V(x_i)+2c\sum_{i>j=1}^{N}\delta(x_i-x_j),
\label{eq:hamiltonian}
\end{equation}
where
\begin{equation}
V(x_i)=\left\{
\begin{array}{cc}
0\;\;\;\;\;\;|x|<L/2,\\
V_0^2\;\;\;\;\;|x|>L/2.
\end{array}
\right.
\nonumber 
\end{equation}
eq. (\ref{eq:hamiltonian}) is the first quantization version of 
Gross-Pitaevski \cite{G-P} equation, which was widely used to study
the Bose-Einstein condensation\cite{BEC} in recent years. 
The model  Hamiltonian (\ref{eq:hamiltonian}) was solved by means of 
Bethe-ansatz method\cite{Li}. 
The logarithm of the Bethe ansatz equation reads
\begin{equation}
\frac{2\pi}{L}{I_j}=k_j
  +\frac{2}{L}\sin^{-1}(\frac{k_j}{V_0})
   +\frac{1}{L}\sum_{l\neq{j}}
     \left[\tan^{-1}(\frac{k_j+k_l}{c})
      +\tan^{-1}(\frac{k_j-k_l}{c})\right],
\label{eq:seculareq}
\end{equation}
where the integers $I_j$ play the role of the quantum numbers.
Equation (\ref{eq:seculareq}) is the secular
equation to determine  the spectrum.
Moreover the transcendental equations (\ref{eq:seculareq}) is 
difficult to solve directly. 
Now, we consider the problem in thermodynamic limit: 
$N\gg 1$ and $L\gg 1 $ 
with a fixed concentration $D=N/L$ 
by introducing  a ``smooth" positive-defined density $\bar{\rho}(k)$
describing the distribution of roots and holes\cite{Lowenstein} 
\[
\bar{\rho}(k)=\frac{1}{L}\frac{dI(k)}{dk}.
\]
Treating $k_j$ as continuous variable $k$ and differentiating 
eq(\ref{eq:seculareq}) with respect to $k$, 
we get an integral equation  
\begin{eqnarray}
2\pi\bar{\rho}(k)=1 +\frac{1}{L}\frac{2}{\sqrt{V_0^2-k^2}}
 +\int dk'\bar{\rho}(k')\left[\frac{c}{c^2+(k-k')^2}
  +\frac{c}{c^2+(k+k')^2}\right]
    \nonumber\\
 -\frac{1}{L}\sum_m\left[\frac{c}{c^2+(k-h_m)^2}
   +\frac{c}{c^2+(k+h_m)^2}\right],
\label{eq:dequation}
\end{eqnarray}
where we have used the replacement
$$\lim_{N,L\rightarrow\infty}\frac{1}{L}\sum_{l\neq j}f(k_l)
  =\int\bar{\rho}(k)f(k)dk-\frac{1}{L}\sum_m f(h_m),$$
in the thermodynamics limit. The summation in the right hand side
runs over ``holes'' (including $k_j$) which can be written formally
as an integral $\int\rho_h(k)f(k)dk$ with 
$\rho_h(k)=(1/L)\sum_m\delta(k-h_m)$. 
Furthermore, eq. (\ref{eq:dequation}) 
is write out as an integral equation for the density of holes
$\rho_h(k)$ and the density of roots $\rho(k)=\bar{\rho}(k)-\rho_h(k)$,
\begin{equation}
2\pi(\rho+\rho_h)=1
 +\frac{1}{L}\frac{2}{\sqrt{V_0^2-k^2}}
  +\int dk'\rho(k')[\frac{c}{c^2+(k-k')^2}
    +\frac{c}{c^2+(k+k')^2}],
\label{eq:densityeq}
\end{equation}

As the Bethe ansatz solution is obtained for the case 
of bounded states (i.e.
${\rm Im} \kappa_j=0$), so the range of the integration is 
$[-V_0, V_0]$. 
In terms of the distribution function of roots, we can write 
out the energy 
per particle
\begin{equation}
E/N=D^{-1}\int\rho(k)k^2dk,
\label{eq:energy}
\end{equation}
where
\begin{equation}
D=N/L=\int\rho(k)dk.
\label{eq:density}
\end{equation}
On the basis of the strategy of ref.\cite{YYang}, the total 
entropy of the system is obtained  
\begin{equation}
S/N=D^{-1}\int[(\rho+\rho_h)\ln(\rho+\rho_h)
     -\rho\ln\rho-\rho_h\ln\rho_h]dk.
\label{eq:entropy}
\end{equation}
where the Boltzmann constant is put to unit. 

In the thermal equilibrium, the free energy 
$\Omega=(E-TS-\mu{N})$ should be in minimum.
Writing
\begin{equation}
\frac{\rho_h(k)}{\rho(k)}=\exp[\epsilon(k)/T],
\label{eq:epsilondef}
\end{equation}
we obtain from the minimizing requirements $\delta\Omega=0$ 
the following equations,  
\begin{equation}
\epsilon(k)=-\mu+k^2
  -\frac{T}{2\pi}\int[\frac{c}{c^2+(k-k')^2}
   +\frac{c}{c^2+(k+k')^2}]\ln(1+e^{-\epsilon(k')/T})dk'.
\label{eq:epsiloneq}
\end{equation}
Eq. (\ref{eq:densityeq}) is readily written as 
\begin{equation}
2\pi (1+e^{\epsilon/T})\rho(k)=1
   +\frac{1}{L}\frac{2}{\sqrt{V_0^2-k^2}}
    +\int dk'\rho(k')[\frac{c}{c^2+(k-k')^2}
     +\frac{c}{c^2+(k+k')^2}].
\label{eq:thedensityeq}
\end{equation}
Principally, eq. (\ref{eq:epsiloneq}) can be solved by iteration and 
then eq. (\ref{eq:densityeq}) can be  
a Fredholm type equation for $\rho(k)$.

We would like to mention some points about the parameter $\mu$. 
If minimizing the Helmholtz free energy $F=E-TS$ under the condition
that the concentration $D$ in (\ref{eq:density}) is a constant, 
one will have a 
Lagrangian multiplier. The multiplier function is just the same as 
the chemical potential $\mu$ when considering a grand assemble.
So both procedures are equivalent. 

Multiplying eq. (\ref{eq:epsiloneq}) with $\rho{D^{-1}}$ and
integrating over $k$, we obtain
\begin{equation}
\mu=D^{-1}\int(k^2-\epsilon)\rho{dk}
  -\frac{T}{2\pi{D}}\int[2\pi(\rho+\rho_h)-1
   -\frac{1}{L}\frac{2}{\sqrt{V_0^2-k^2}}]\ln(1+e^{-\epsilon/T})dk.
\label{eq:mu}
\end{equation}
The argument $k$ of $\rho$ and $\epsilon$ are always omitted 
in our notions as long as 
it does not bring about confusions. 
With the help of eq. (\ref{eq:epsilondef}), 
the entropy (\ref{eq:entropy})
is rewritten as  
\begin{equation}
\frac{S}{N}=D^{-1}\int[(\rho+\rho_h)\ln(1+e^{-\epsilon/T})+\rho\epsilon/T]dk.
\nonumber 
\end{equation}
The Helmholtz free energy per particle is
\begin{equation}
\frac{F}{N}=\frac{1}{D}\int(k^2-\epsilon)\rho{dk}
 -\frac{T}{D}\int[(\rho+\rho_h)\ln(1+e^{-\epsilon/T})dk.
\label{eq:freenergy}
\end{equation}
Comparison of eq. (\ref{eq:mu}) and eq. (\ref{eq:freenergy}) gives rise to
\begin{equation}
F=\mu{N}-\frac{TL}{2\pi}\int[1+\frac{1}{L}
          \frac{2}{\sqrt{V_0^2-k^2}}]\ln(1+e^{-\epsilon/T})dk.
\label{eq:thefreenergy}
\end{equation}
Thus the free energy will be obtained once the $\epsilon(k)$ is solved 
from eq. (\ref{eq:epsiloneq}). As in thermodynamics
$F=-PL+\mu N$, the pressure is
$P=-(\partial{F}/\partial{L})_T$.
It was shown\cite{GLiL} that if $\epsilon$, $\mu$ are implicit functions
of some thermodynamic quantities $x$ (such as $T$, $L$), the derivative
of eq. (\ref{eq:thefreenergy}) with respect to $x$ is the same 
as the partial derivative of
eq. (\ref{eq:thefreenergy}) with respect to the explicit variable $x$. 
Then it is easy to
calculate the pressure in terms of the $\epsilon$, namely, 
\begin{equation}
P=\frac{T}{2\pi}\int[1+\frac{1}{L}\frac{2}{\sqrt{V_0^2-k^2}}]
   \ln(1+e^{-\epsilon/T})dk.
\label{eq:presure}
\end{equation}
It is formally similiar to the result of ref.\cite{YYang} 
except one more term arising from the boundary condition is involved. 
Like-wisely, the entropy is  
\begin{equation}
S=\frac{L}{2\pi}\int[1+\frac{1}{L}\frac{2}{\sqrt{V_0^2-k^2}}]
   [\ln(1+e^{-\epsilon/T})+\frac{\epsilon/T}{1+e^{\epsilon/T}}]dk.
\label{eq:thentropy}
\end{equation}
The other thermal quantities such as specific heat etc. 
are also obtainable if one is able to solve $\epsilon(k)$ 
from eq. (\ref{eq:epsiloneq}). We will calculated them  in some special cases
in the following sections.

\section{The strong coupling limit}
 
It is difficult to obtain an explicit expression
of $\epsilon(k)$ from eq. (\ref{eq:epsiloneq}) generally. 
However, in some special cases,
we are able to obtain some plausible results. 
In strong coupling limit $c \gg V_0$, 
eq(\ref{eq:epsiloneq}) becomes
\begin{equation}
\epsilon(k)=-\mu+k^2
  -\frac{T}{2\pi}\int\ln[1+e^{(\mu-k'^2)/T}]
    \frac{d}{dk'}[\tan^{-1}(\frac
      {k^\prime-k}{c})+\tan^{-1}(\frac{k'+k}{c})]dk'.
\label{eq:Sepsiloneq}
\end{equation}
Integrating by part under the consideration of approximation
that
$\tan^{-1}(k/c)\simeq{k/c}$ for $c \gg V_0$,
we have 
\begin{equation}
\epsilon=-\mu'+k^2,
\label{eq:Sepsilon}
\end{equation}
where
\begin{equation}
\mu'=\mu+\frac{2}{\pi c}\int\frac{k'^2}{1+e^{(-\mu+k'^2)/T}}dk',
\label{eq:muprime}
\end{equation}
Because of
$\displaystyle\frac{2}{L}\sqrt{V_0^2-k^2}\ll 1$   
and $c \gg V_0$, we obtain up to the first order that
\begin{equation}
2\pi(\rho+\rho_h)=1
 +\frac{1}{L}\frac{2}{\sqrt{V_0^2-k^2}}
  +\frac{4V_0}{c}.
\label{eq:Srhohole}
\end{equation}
From eq. (\ref{eq:epsilondef}),(\ref{eq:Sepsilon}) 
and (\ref{eq:Srhohole}), we obtain an 
analytic expression of $\rho(k)$:
\begin{equation}
2\pi\rho(k)=[1+\frac{1}{L}\frac{2}{\sqrt{V_0^2-k^2}}
     +\frac{4V_0}{c}][1+e^{(-\mu'+k^2)/T}]^{-1}.
\label{eq:Srho}
\end{equation}
Obviously, the $\rho(k)$ is a Fermi-like distribution. 
When $c, L\rightarrow\infty$,
\begin{equation}
2\pi\rho(k)=\frac{1}{1+e^{(-\mu+k^2)/T}},
\label{eq:Fermi}
\end{equation}
which is just the distribution of free Fermi gas. 
The chemical potential $\mu$ should
be positive-definite for the positive-definite density of roots.

As all particles are 
bounded in the potential well, i.e. $Max(k)\sim V_0$,
$\rho(k)$ should
vanish almost for $k>V_0^{1/2}$. This requirement together 
with eq. (\ref{eq:Srho}) gives
\[
(V_0^2-\mu')/T\gg 1.
\] 
The system being thermal equilibrium exhibits a simple 
dependence on the large momentum cut-off $T_0$
($T_0=V_0^2-\mu'$), due to boundary effects. 
Therefore, the system can be in a state of thermal equilibrium
only when $T\ll T_0$. Otherwise some particles
may overcome the potential energy at the boundary and escape out of the well.
Substituting the obtained $\epsilon$ into eq. (\ref{eq:thefreenergy}),
we obtain
\begin{equation}
F=\mu N-\frac{TL}{2\pi}\int\ln[1+e^{(\mu'-k^2)/T}]
   \frac{d}{dk}[k+\frac{2}{L}\sin^{-1}(\frac{k}{V_0})]dk.
\end{equation}
As $k\ll V_0$,  we can replace $\sin^{-1}(k/V_0)$ by $k/V_0$, then
\begin{equation}
F=\mu N-\frac{2L}{\pi}(1+\frac{2}{LV_0})
   \int_0^{V_0}\frac{k^2}{1+e^{(-\mu'+k^2)/T}}dk.
\label{eq:Sfreenergy}
\end{equation}

In the low-temperature condition, the free energy becomes
\begin{equation}
F=\mu N-\frac{2L}{\pi}(1+\frac{2}{LV_0})(\frac{1}{3}\mu'^{3/2}
        +\frac{T^2\pi^2}{24\mu'^{1/2}}),
\label{eq:Sthefreenergy}
\end{equation}
where
\begin{equation}
\mu'=\mu+\frac{1}{c}(\frac{4}{3\pi}\mu^{3/2}
      +\frac{T^2\pi}{6\mu^{1/2}}).
\end{equation}
The $\mu'$ is regarded as a mandation of chemical potential 
according to eq. (\ref{eq:Srho}).
However, we are not able to get an explicit result for 
the specific heat by partial derivative of eq. (\ref{eq:Sthefreenergy}), 
because the chemical potential $\mu$ might be temperature dependent. 
In order to observe 
some properties of specific heat at low temperature, 
we let $c \rightarrow\infty$ and let 
$\mu_0$ denote chemical potential at zero temperature. 
The $\mu_0$ is determined by
\begin{equation}
D=\frac{1}{2\pi}\int_{-\sqrt{\mu_0}}^{\sqrt{\mu_0}}
  (1+\frac{1}{L}\frac{2}{\sqrt{V_0^2-k^2}})dk.
\end{equation}
Considering $\lim_{T\rightarrow 0}\mu(T)/\mu_0=1$, we have
\begin{equation}
F=\mu_0 N-\frac{2L}{\pi}(1+\frac{2}{LV_0})
   (\frac{1}{3}\mu_0^{3/2}+\frac{T^2\pi^2}
    {24\mu_0^{1/2}}).
\label{eq:SF}
\end{equation}
We find that the specific heat at low temperature is Fermi-liquid like
\begin{equation}
C_V=\frac{\pi L}{6{\mu_0}^{1/2}}(1+\frac{2}{LV_0})T.
\label{eq:Sspecific}
\end{equation}
Thus the interaction between the particles 
plays an important role to their statistical properties
though the system we considered is a boson system. 
This is a model belong to the class of Haldane's exclusion 
statistics\cite{Haldane2}. 

\section{The weak coupling limit}

Obviously, eq. (\ref{eq:epsiloneq}) can be written as
\begin{equation}
\epsilon=-\mu+k^2-\frac{T}{2\pi}\int{e}^{-c|\omega|}e^{ik\omega}
  \cos(k'\omega)\ln(1+e^{-\epsilon/T})dk'd\omega.
\label{eq:Wepsilon}
\end{equation}
Because in weak coupling limit $c \ll 1$, we have
\begin{equation}
\epsilon(k)=-\mu+k^2-T\ln(1+e^{-\epsilon/T})-f(k,c),
\label{eq:Wthepsilon}
\end{equation}
where
\[
f(k,c)=\frac{T}{2\pi}
   \sum_{n=1}^{\infty}\int\frac
     {(-1)^nc^n|\omega|^n}{n!}e^{ik\omega}\cos(k'\omega)\ln(1
       +e^{(\mu-k^2)/T})dk'd\omega.
\]
This leads to
\begin{equation}
e^{-\epsilon/T}=[e^{(-\mu+k^2-f(k,c))/T}-1]^{-1}.
\label{eq:Wexpepsilon}
\end{equation}
Like-wisely, eq. (\ref{eq:densityeq}) gives rise to
\begin{equation}
2\pi\rho_h(k)=1+\frac{1}{L}\frac{2}{\sqrt{V_0^2-k^2}}+g(k,c),
\label{eq:Whole}
\end{equation}
where
\[
g(k,c)
  =\sum_{n=1}^\infty\int\frac{(-1)^nc^n|\omega|^n}
    {n!}e^{ik\omega}\cos(k'\omega)dk'd\omega.
\] 
With the help of (\ref{eq:Wexpepsilon}), 
we obtain the distribution function of roots
\begin{equation}
2\pi\rho(k)=[1+\frac{1}{L}\frac{2}{\sqrt{V_0^2-k^2}}
  +g(k,c)]\frac{1}{e^{[-\mu+k^2-f(k,c)]/T}-1}.
\label{eq:Wrho}
\end{equation}
Physically, it represents the distribution of quasi-momenta of the 
system as a collection.
Because the density of root should be positive-definite, 
$\mu+f(k,c)$ must be always smaller
than the corresponding $k^2$, particularly, 
$\mu+f(0,c)\leq 0$.
So eq. (\ref{eq:Wrho}) is boson-like distribution. 
If considering the boundary effects so that 
$\rho(V_0)\simeq 0$ we have 
$V_0^2-\mu-f(V_0,c)\gg{T}$.
There is a large-momentum cut-off $T_0=V_0^2-f(V_0,c)-\mu$ such that 
the system can be in a thermal equilibrium only when $T\ll T_0$.

Now we consider the free energy, and  only take 
account of the leading terms for small $c$,
\begin{equation}
2\pi\rho(k)=[1+\frac{1}{L}\frac{2}{\sqrt{V_0^2-k^2}}]
  \frac{1}{e^{(k^2-\mu)/T}-1}.
\label{eq:Wtherho}
\end{equation}
The free energy (\ref{eq:thefreenergy}) becomes
\begin{equation}
F=\mu{N}+T\frac{L}{2\pi}\int[1
   +\frac{1}{L}\frac{2}{\sqrt{V_0^2-k^2}}]
     \ln[1-e^{(\mu-k^2)/T}]dk.
\end{equation}
Since $k\ll{V_0}$, $\sin^{-1}(k/V_0)\simeq{k}/V_0$, we have  
\begin{equation}
F=\mu{N}-\frac{L}{\pi}(1+\frac{2}{LV_0})
     \int\frac{k^2}{e^{(k^2-\mu)/T}-1}dk.
\label{eq:Wfreenergy}
\end{equation}
As for free bosons (i.e., $c=0$), the chemical potential should be 
nonpositive-definite 
and smaller than the energy of any particles. 
Moreover, for a fixed concentration ($D=N/L$), 
it is a function of temperature decreasing as the temperature increases 
according to eq. (\ref{eq:Wtherho}). 
Hence when temperature goes to zero, the chemical
potential will approach to zero from negative value.

At low temperature, eq. (\ref{eq:Wfreenergy}) becomes
\begin{equation}
F=\mu{N}-\frac{LT^{3/2}}{2\sqrt{\pi}}(1+\frac{2}{LV_0})Li_{3/2}(e^{\mu/T}),
\label{eq:Wthefreenegy}
\end{equation}
where $Li_n(z)$ is polylogarithm function with $Li_{3/2}(1)=\zeta(3/2)$ and
$\zeta(x)$ is Riemann's Zeta function. Since $Li_n(z)$ can be expanded into
series of $z$ and the chemical potential $\mu$ should be zero at zero 
temperature, we neglect the $Li_{3/2}(e^{\mu/T})$'s dependence on $T$. 
Then the entropy has the form
\begin{equation}
S=\frac{3LT^{1/2}}{4\sqrt{\pi}}(1+\frac{2}{LV_0})Li_{3/2}(e^{\mu/T}), 
\label{eq:Wthentropy}
\end{equation}
and the specific heat is proportional to $T^{1/2}$ 
which exhibits  bose-gas like behaviors.

\section{Conclusions and remarks}

In the above, we discussed the thermodynamics of bosons in 
a one-dimensional potential well on the basis of the exact solution of 
the model. 
Using the strategy of Yang and Yang\cite{YYang},
we studied the
general thermodynamic properties of the system. We considered
the problem in strong coupling limit and found that the behavior
of the system at low temperature is Fermi liquid like even though it is
a boson system. Therefore the interaction plays an important role. 
Meanwhile we obtained the specific heat which is
linearly dependent on the  temperature $T$. 
For the weak coupling limit, we found that the 
system behaves like free boson gas at low temperature. 

\section*{Acknowledgment}

This work is supported by NSFC No.19975040 and EYF98.
YQL thanks T.-L. Ho for helpful discussions and
acknowledges the support by AvH-Stiftung in completing part of the work.

\end{document}